\documentclass[twocolumn,prb,superscriptaddress]{revtex4-2}
\usepackage{graphicx,amssymb,soul,color,amsmath,bm}
\usepackage{epstopdf,hyperref}
\usepackage{braket,dsfont,nicefrac}
\usepackage[mathcal]{euscript}
\usepackage[normalem]{ulem}
\usepackage[utf8]{inputenc}

\hypersetup{colorlinks=true,citecolor=blue,linkcolor=magenta}

\sethlcolor{yellow}
\allowdisplaybreaks

\begin{document}

\title{The fate of pairing and spin-charge separation in the presence of long-range antiferromagnetism}

\author{Luhang Yang}
\affiliation{Department of Physics, Northeastern University, Boston, Massachusetts 02115, USA}

\author{Ignacio Hamad}
\affiliation{Instituto de F\'isica Rosario (CONICET) and Facultad de Ciencias Exactas, Ingenier\'ia y Agrimensura, Universidad Nacional de Rosario, Rosario, Argentina}

\author{Luis O. Manuel}
\affiliation{Instituto de F\'isica Rosario (CONICET) and Facultad de Ciencias Exactas, Ingenier\'ia y Agrimensura, Universidad Nacional de Rosario, Rosario, Argentina}

\author{Adrian E. Feiguin}
\affiliation{Department of Physics, Northeastern University, Boston, Massachusetts 02115, USA}

\begin{abstract}
We present a numerical study of competing orders in the 1D $t$-$J$ model with long-range RKKY-like staggered spin interactions. By circumventing the constraints imposed by Mermin-Wagner's theorem, this Hamiltonian can realize long-range N\'eel order at half-filling. We determine the full phase diagram as a function of the exchange and particle density using the density matrix renormalization group (DMRG) method. We show that pairing is disfavored and the AFM insulator and metallic phases are separated by a broad regime with phase segregation, before spin-charge separation re-emerges at low densities. Upon doping, interactions induce a confining potential that binds holons and spinons into full fledged fermionic quasi-particles in a range of parameters and densities. We numerically calculate the photoemission spectrum of the model, showing the appearance of a coherent quasi-particle band splitting away from the holon-spinon continuum with a width determined by $J$ that survives at finite doping. Comparison with analytical results using the self-consistent Born approximation (SCBA) and by solving the spinon-holon problem offer insight into the internal structure of the quasi-particles and help us explain the different features in the spectrum. We discuss how this simple toy-model can teach us about the phenomenology of its higher dimensional counterpart.

\end{abstract}
\maketitle

\section{Introduction}

Understanding the properties of doped antiferromagnets has been a topic of great theoretical interest for the past few decades\cite{Dagotto1994,Lee2006}. This is motivated by the lack of a universal theory of high-temperature superconductors that can explain the mechanisms behind the formation of Cooper pairs in this kind of materials, where strong electronic correlations are assumed to play a dominant role. Most of the research in this area has been focused on the study of paradigmatic simple model Hamiltonians that are supposed to capture all the basic ingredients for high temperature superconductivity such as the Hubbard and $t-J$ models and variations of them\cite{Dagotto1994,Imada1998,ScalapinoReview,qin2021hubbard,SimonsCollab2015,SimonsCollab2020}. In this context, much effort has been devoted to their low-dimensional versions\cite{frahm90,penc97,jeckelmann00,benthien04,carmelo04,carmelo06,carmelo08,Kohno2010,SIG10a,pereira10,essler10,SIG10b,SIG12,seabra14,essler15,tiegel16,veness16,Yang2016cpt,Wang2015}. Particularly, in one dimension the physics of these systems can be universally described in the framework of Luttinger liquid (LL) theory\cite{Tomonaga1950,Luttinger1963,mattis_lieb_1965,Haldane1981,Gogolin,GiamarchiBook}: the natural excitations in 1D are described in terms of spin and charge excitations that propagate coherently with different velocities and are characterized by distinct energy scales, leading to the concept of spin-charge separation. The spectrum of a spin-full LL is determined by a convolution of the spin and charge spectra, which leads to a continuum without well defined Landau quasi-particles and Fermi-edge singularities instead of quasi-particle peaks\cite{Essler2003,Carmelo2021}. Interestingly, the Hubbard model in 1D admits an exact solution in terms of the Bethe ansatz \cite{Lieb1968,LIEB2003,EsslerBook}, and the $t-J$ model also realizes an exactly soluble ``supersymmetric'' point at $J/t=2$\cite{Bares1990,Bares1991,Sarkar1991,Essler1992,Moreno2013}, allowing one to infer information about the nature of the excitations.  

\begin{figure}
	\centering
	\includegraphics[width=0.45\textwidth]{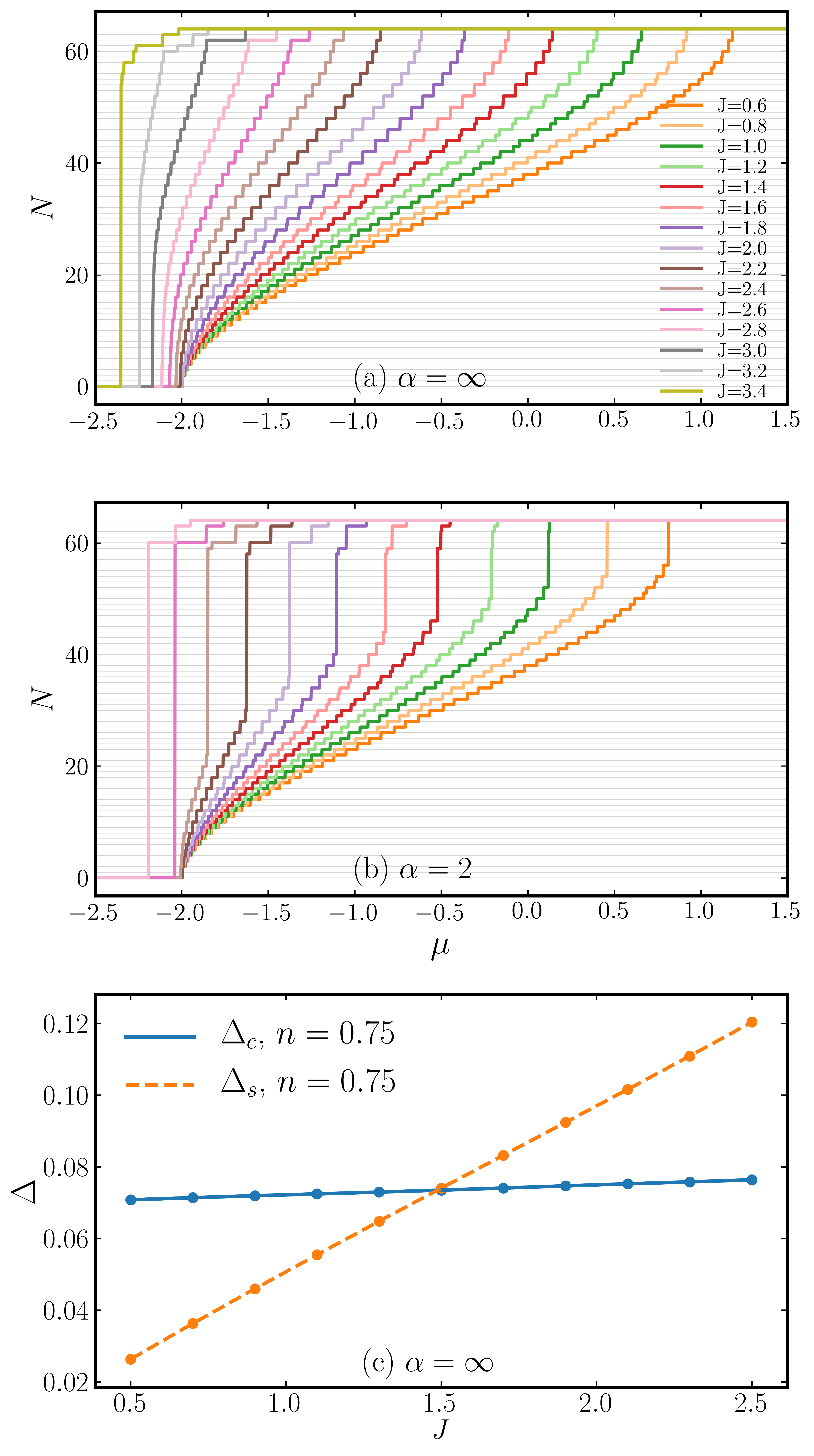}
	\caption{Particle number as a function of chemical potential for the (a) conventional $t-J$ model ($\alpha=\infty$) and (b) $\alpha=2$ for several values of $J$ in descending order from left to right.(c) Charge and spin gaps for a conventional $t-J$ chain with $L=64$ sites.}
	\label{fig:N_mu}
\end{figure}

An important difference between one and two dimensions is established by the Mermin-Wagner theorem\cite{Mermin1966,Halperin2019}: at zero temperature, gapless one-dimensional local Hamiltonians cannot realize long-range order, while two dimensional systems can display spontaneous symmetry breaking. At half-filling, where the physics can be more easily understood in the context of the Heisenberg model, the ground state of a 1D chain is a spin-liquid with algebraically decaying correlations and domain-wall-like spin excitations (spinons) that carry spin $S=1/2$. On the other hand, the ground state in two dimensions displays N\'eel order, and excitations are magnons that condense into Goldstone modes.\cite{Zhitomirsky2013}

Besides the omnipresent question concerning the role of antiferromagnetism as a glue for pairing, a more basic and fundamental one has also remained central to the problem: is it possible for spin-charge separation to survive in two dimensions? \cite{Putikka1994,Chen1994,Haas1996a,Rice1997,Martins1999,Tremblay1995,Poilblanc1995,Arrigoni1999,Vishwanath2001,Anderson2000,Brunner2001,Martins2000a,Martins2000b,Mishchenko2001,Andreas2004,Bonca2007,Moritz2009,Grusdt2018scipost,Grusdt2019microscopic,Bohrdt2020}. Alternatively, one can postulate the opposite question: What is the fate of spin charge separation in the presence of long-range antiferromagnetic order? The spin-charge separation phenomenon is usually considered as a manifestation of 1D physics, and whether it exists in higher dimensions is a topic of debate, especially in the context of understanding high-Tc superconductivity\cite{Anderson2000} and recent experiments in cold-atom systems\cite{Jordens2008,Esslinger2010,Endres2011,Georgescu2014,Hart2015,Duarte2015,Boll2016,Cheuk2016,Brown2017,Hilker2017,Mazurenko2017,Gross2017,Scherg2018,Salomon2019,Nichols2019,Chiu2019,Koepsell2019}. Over the past decades it has become quite clear that a definitive answer to these questions can only be obtained numerically. Unfortunately, quantum Monte Carlo (QMC) has not been able to provide evidence since calculations are carried out at finite temperature and with the use of difficult to control analytic continuation\cite{Bulut1994,Preuss1995,Preuss1997,Grober2000}. At the same time, the success of the density matrix renormalization group method (DMRG) \cite{White1992,White1993} and tensor networks has only partially extrapolated to two dimensions\cite{Corboz2011,Corboz2014,Corboz2016,SimonsCollab2017}. 

\begin{figure}
	\centering
	\includegraphics[width=0.43\textwidth]{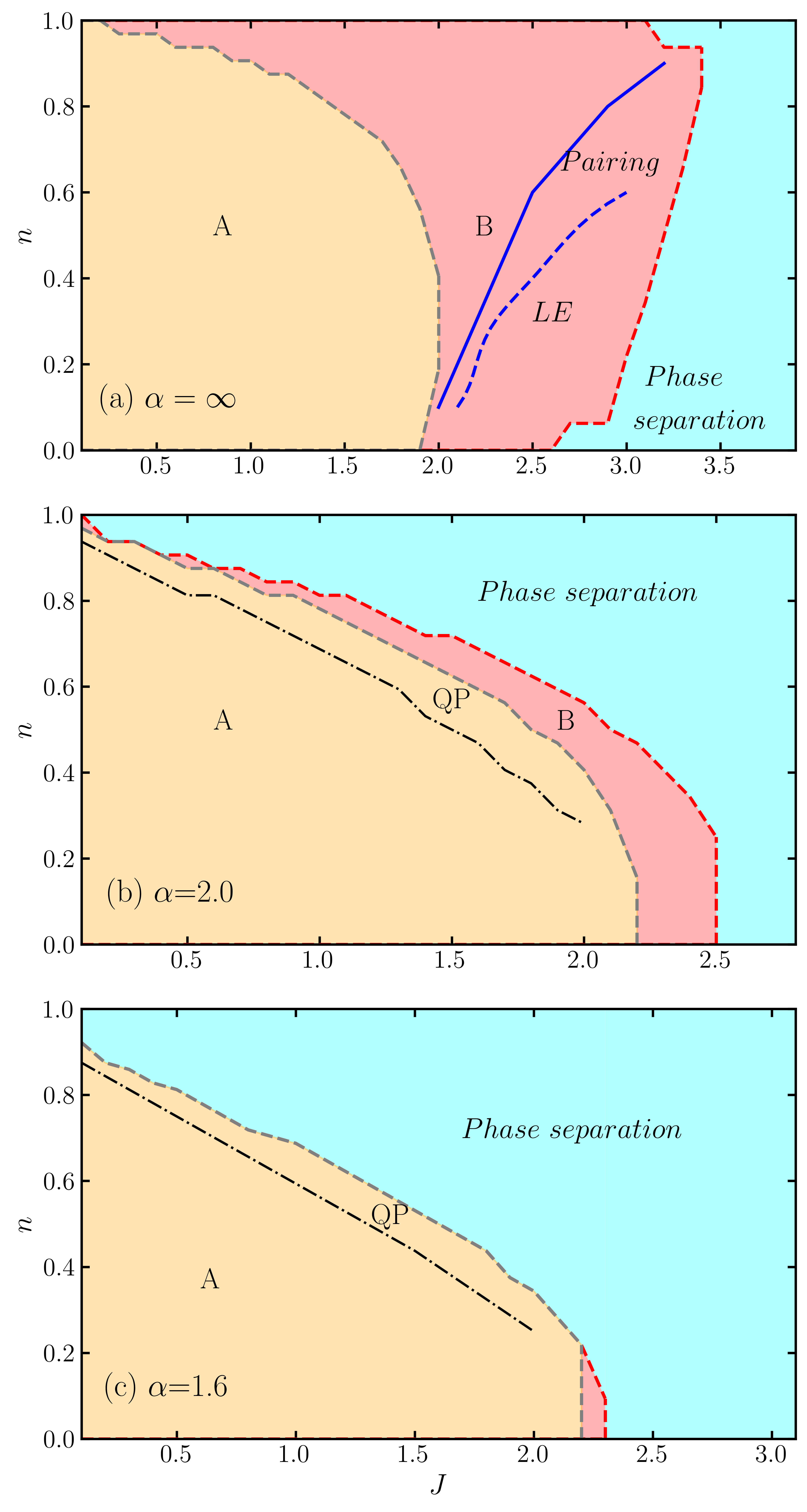}
	\caption{(a) Phase diagram of the conventional $t-J$ chain. Regions A and B represent regimes where the occupation changes by $\Delta N=1$ and $\Delta N=2$ with varying the chemical potential, respectively. The solid and dashed blue curve are the phase boundaries of the paired and Luther-Emery phase, respectively. (b)-(c) Phase diagram for the $t-J$ chain with long range spin interactions and (b) $\alpha = 2$ and (c) $\alpha=1.6$. The dotted-dashed line indicates the boundary between phase separation and fermionic quasiparticles (see text). All results are for a chain with $L=64$ sites.}
	\label{fig:phase_diagram}
\end{figure}

One possible avenue to circumvent these hurdles and study the dimensional crossover consists of introducing long-range interactions in one-dimension since\cite{Yusuf2004,Laflorencie2005,Sandvik2010,Tang2015,Yang2021,Yang2020deconfined,Yang2021ladder}: (i) they effectively increase the dimensionality of the problem through all-to-all interactions; (ii) overcome the limitations of Mermin-Wagner theorem allowing one to probe for true long range order and spontaneous symmetry breaking; (iii) they offer a relatively simple and intuitive playground where to test for higher dimensional physics within the reach of powerful numerical techniques such as the DMRG method. 


In this work, we focus on understanding the role of long range interactions in a doped one-dimensional antiferromagnet using an extended $t-J$ model with RKKY-like AFM long-range interactions:

\begin{eqnarray}
H&=&H_{t-J}+H_{RKKY} \\
    H_{t-J} & = & -t\sum_{i\sigma}(c_{i,\sigma}^\dagger c_{i+1,\sigma} + h.c.) \\ \nonumber
    & + & J\sum_{i}(\vec{S}_i\cdot \vec{S}_{i+1} - \cfrac{1}{4} n_in_{i+1}) \\
    H_{RKKY} &=& \lambda\sum_{i,j>i+1}\cfrac{ (-1)^{j-i+1}}{|j-i|^\alpha}(\vec{S}_i\cdot \vec{S}_{j}),
\label{hami}
\end{eqnarray}
where the operator  $c^\dagger_{i\sigma}$ creates an electron on site $i$ along the chain with spin $\sigma=\uparrow,\downarrow$, $n_{i}$ is the electron number operator, $\vec{S}$ represents spin $S=1/2$ operators. The constants $J$ and $\lambda$ parametrize the magnitude of the spin exchange and RKKY interactions that decay as a power law with exponent $\alpha$. In the rest of the paper and for simplicity, we focus on the case $\lambda=J$ and we study finite chains of length $L$. The $t-J$ model describes the low-energy physics of the Hubbard model when the Coulomb repulsion is very large compared to the hopping constant $t$, that we take as our unit of energy. In this context, a constraint forbidding double-occupancy is implicitly assumed. 

\begin{figure}
	\centering
	\includegraphics[width=0.45\textwidth]{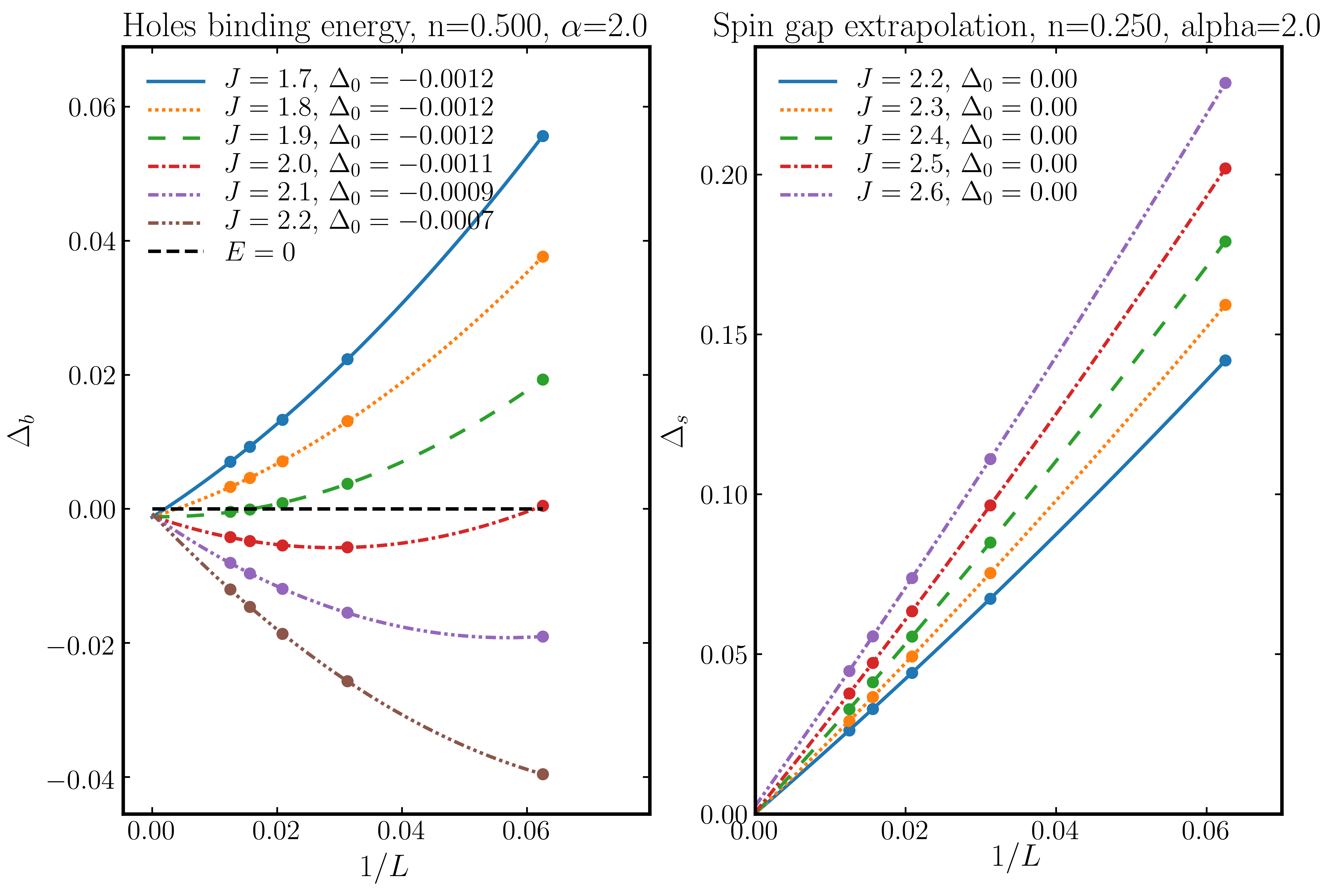}
	\caption{Left panel: spin gap extrapolated to thermodynamic limit with second order polynomial fit. Calculated by DMRG for density $n=0.25$ and $\alpha=2$, $J$ ranges from metallic phase to the phase separation boundary. Right panel: binding energy between two holes, extrapolated to the thermodynamic limit with second order polynomial fit for $n=0.5$, $\alpha=2$. $J$ varies through region B in phase diagram Fig.\ref{fig:phase_diagram}(b).}
	\label{fig:spin_gap}
\end{figure}

The quantum phase diagram of conventional $t-J$ model in 1D has been extensively studied\cite{Ogata1991,Moreno2011}. At half-filling, this Hamiltonian reduces to the one-dimensional Heisenberg chain. Upon doping and for large $J$, spins prefer to form antiferromagnetic domains and clump together phase separating into electron-rich and hole-rich regions. In the metallic phase with $J/t < 2$, the low energy physics can be well described in terms of Luttinger liquid theory.
In this phase, the low energy excitations are holons carrying charge with characteristic velocity $v_h$ and a bandwidth determined by the hopping $t$, and spinons carrying spins with velocity $v_s$ and a bandwidth proportional to $J$.

In addition to the Luttinger liquid metallic phase, in the intermediate $J/t$ range the $t-J$ model exhibits a Luther-Emery regime at low densities with a spin gap and dominant pairing correlations and a superconducting phase at high densities between the metallic LL phase and phase separation (See Fig.\ref{fig:phase_diagram}(a)).

One can in principle assume that the origin of the RKKY term can be the proximity to a two dimensional layer with long range antiferromagnetic order\cite{Grusdt2018scipost}. Notice that the sign of the interactions alternates between antiferro and ferromagnetic depending on the sublattice and it enhances the tendency  of the spins to antiferromagnetically align. At half-filling, this translates into a regime with spontaneous symmetry breaking and long range order for $\alpha$ small enough $\alpha < 2.2$\cite{Laflorencie2005}. While the elementary excitations of the Heisenberg chain are deconfined domain-wall-like spinons, an effective confining potential emerges as a consequence of the long-range interactions that binds spinons together to form coherent magnon-like gapless excitations above the antiferromagnetic ground state.  In this work, we aim at describing and understanding how similar effects can alter the spin-charge separation picture and induce confinement between spinons and holons such that they bind forming composite quasi-particle states that carry both, spin and charge degrees of freedom. Moreover, we observe that with these modifications to the $t-J$ model, excitations will no longer display a linear dispersion at the Fermi level and LL theory will not apply in its conventional formulation\cite{Eder1997}.

This paper is organized as follow: firstly we study the quantum phase diagram for $t-J$ chain with long range spin couplings and show the dominant orders in different phases in sec.\ref{sec:phase_diagram}. In sec.\ref{sec:polaron} we discuss the stability of composite quasi-particles using energetic considerations. We support this evidence by means of numerical and analytical calculations of the spectral function for a single hole in sec.\ref{sec:spectrum-hole} using DMRG, the self-consistent Born approximation (SCBA) and by solving the spinon-holon problem. We extend these considerations to finite doping in sec.\ref{sec:spectrum}. We finally close with a summary and discussion of the results.


\begin{figure}
	\centering
	\includegraphics[width=0.45\textwidth]{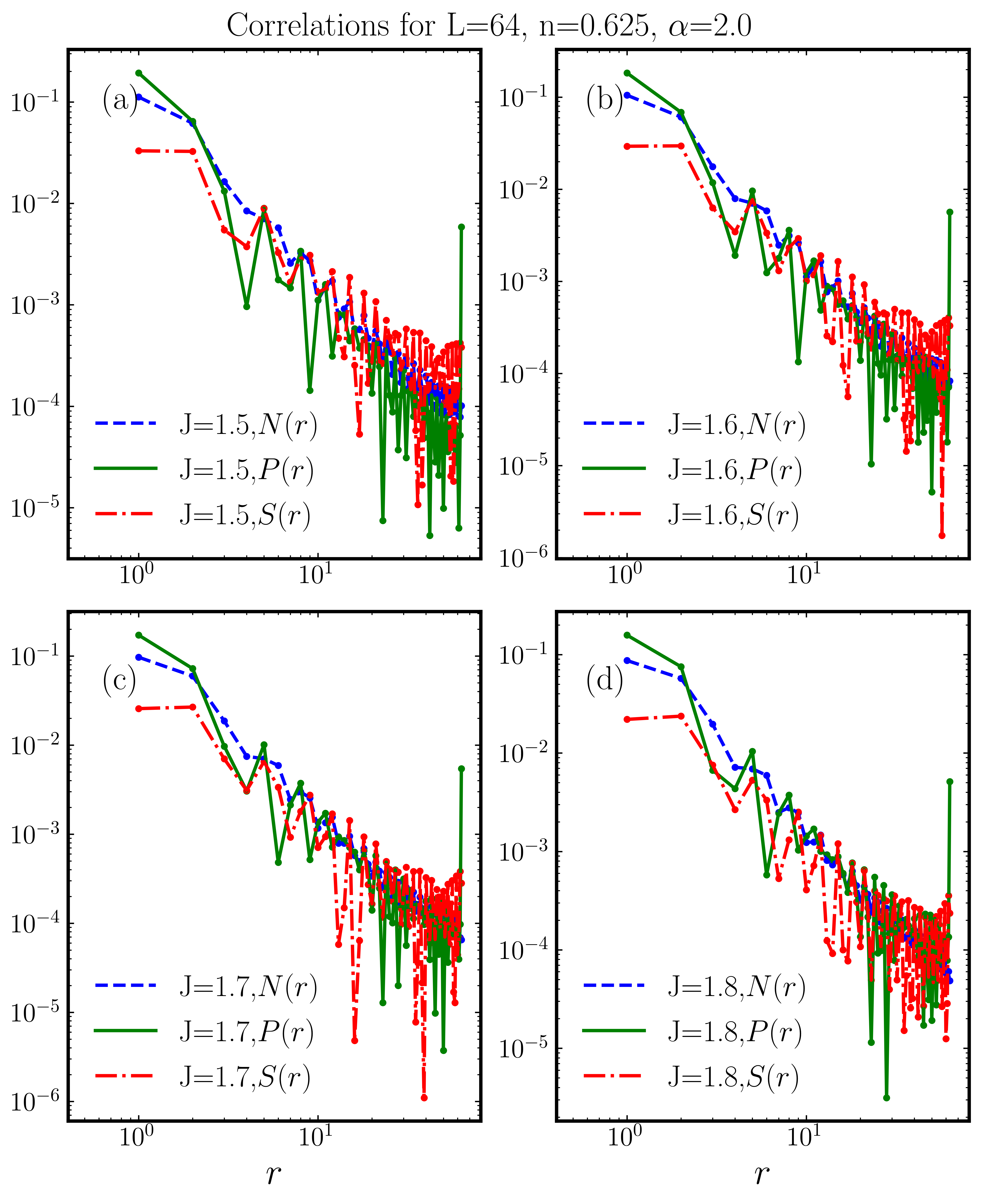}
	\caption{Singlet pair-pair, density-density, and spin-spin  correlations in real space for $L=64$, $n=0.625$, and $\alpha=2$. $J$ varies through region B in phase diagram Fig.\ref{fig:phase_diagram}(b)}
	\label{fig:correlations_4figs}
\end{figure}

\begin{figure}
	\centering
	\includegraphics[width=0.35\textwidth]{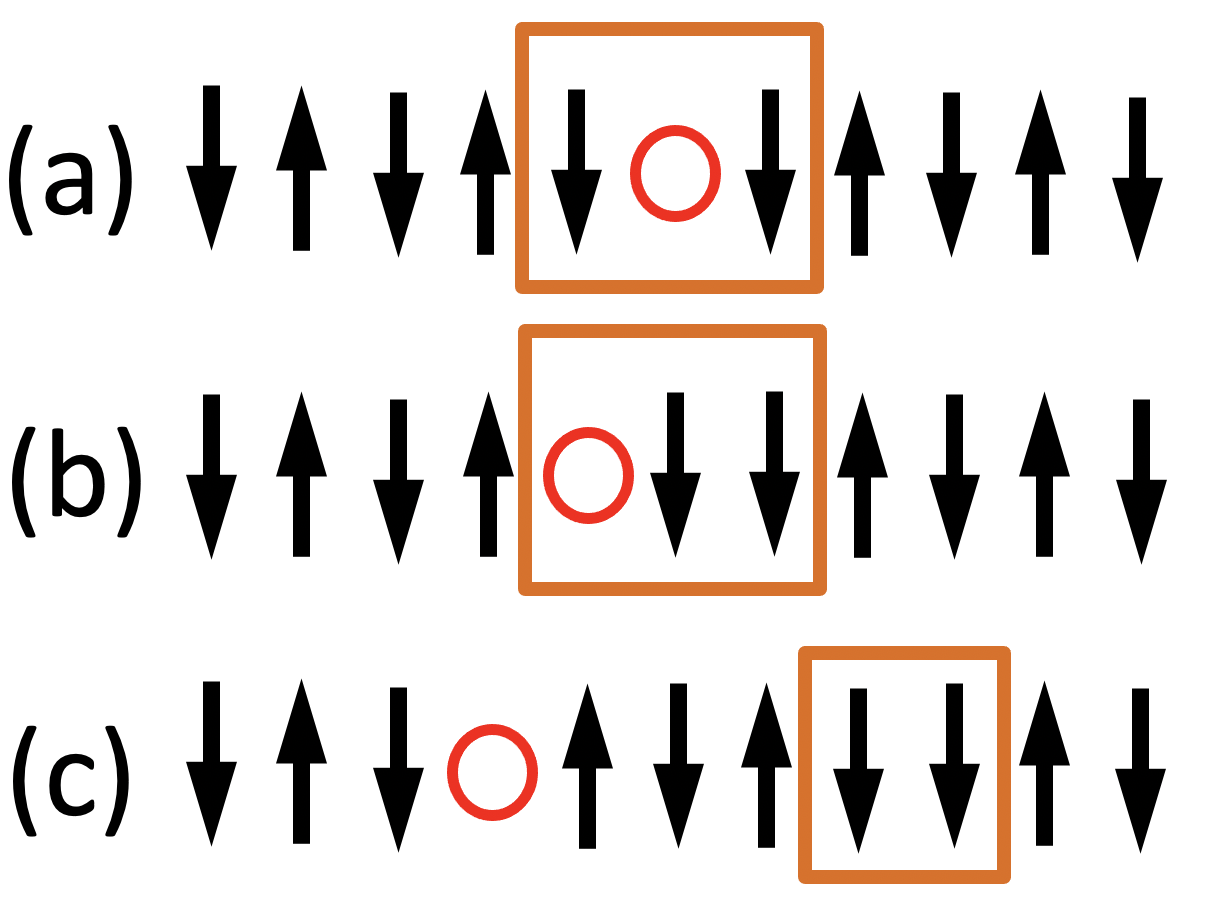}
	\caption{Cartoon picture in the Ising limit illustrating: (a) A confined holon and spinon pair; (b) a spinon a distance $r=1$ from the holon, still forming a bound pair; (c) deconfined spinon and holon. The box highlights the position of the spin domain wall. }
	\label{fig:polaron}
\end{figure}

\section{Phase diagram}
\label{sec:phase_diagram}

In order to determine the quantum phase diagram of the model, we use the DMRG method with open boundary conditions to calculate the ground state energies $E_0(J,N)$ by varying $J$ in steps of 0.1 and changing the total particle number $N$ between 0 and $L$, where $L$ is the length of the chain. In all calculations we chose the bond dimension such that the truncation error is always below $10^{-6}$. Interestingly, despite the long range interaction, we observe that the entanglement does not grow dramatically, allowing us to maintain all sources of error under control (basically, by using enough DMRG states). We use the Maxwell construction to obtain the $N$ versus chemical potential $\mu$ curves and determine the stable ground state densities, as shown in Fig.\ref{fig:N_mu}. For validation, we include results for the conventional $t-J$ chain of length $L=64$. One can distinguish three different behaviors, summarized in the phase diagram  Fig.\ref{fig:phase_diagram}(a): (i) a metallic phase where the occupation changes in steps of one particle ($\Delta N=1$) labeled as ``A''; (ii) a region ``B'' where it changes in steps of two ($\Delta N=2$); and finally (iii) the occupation abruptly jumps between an intermediate value and $n=N/L=1$.  This sudden change is associated to phase segregation: for large $J$ the system splits between hole-rich regions and domains with density $n=1$ and AFM correlations. The steps $\Delta N=2$ can be interpreted as an indication of pairing. However, it turns out that an alternative explanation is possible: due to spin-charge separation, the creation of a hole translates into the excitation of both a spinon and a holon that, as we mentioned, have characteristic velocities $v_s$ and $v_h$. This means that in the regime with $v_s > v_h$ it is energetically more favorable to create two holons without exciting any spinons, rather than a single holon and a spinon\cite{schulz1994,Moreno2013}. This is expected to occur for large enough $J$, which is precisely where this is observed in the phase diagram. To support this argument, we define the singlet-triplet spin gap:
\[
\Delta_s=E(N,S^z=1)-E(N,S^z=0) 
\]
and the charge gap:
\begin{equation}
\begin{split}
& \Delta_c = E(N+1,S^z=1/2)+E(N-1,S^z=1/2) \\
& \ \ \ - 2E(N,S^z=0).
\end{split}
\end{equation}
In Fig.\ref{fig:N_mu}(c) we show both quantities for $L=64$ and density $n=N/L=0.75$. As one can see, the charge gap has a very weak dependence on $J$ and is essentially determined by the level spacing $\Delta_c \sim 4t/L=0.0625$. Both gaps extrapolate to zero in the $L\rightarrow \infty$ limit (not shown) but in finite systems they display a crossing at precisely the value of $J$ where the steps change from $\Delta N=1$ to $\Delta N=2$.
Hence, this is a finite-size effect since both spinons and holons are gapless in the thermodynamic limit. However, it is a feature that should remain observable in finite chains and can help us as a guide in our search for pairing, since that should also manifest itself as steps $\Delta N=2$ as well. 
As a matter of fact, pairing is known to be stable in this regime \cite{Moreno2011}, as shown in the same figure. There are two distinct paired phases: a gapless one with algebraically decaying but dominant pair-pair correlations, and a spin-gapped Luther-Emery phase.  

\begin{figure}
	\centering
	\includegraphics[width=0.45\textwidth]{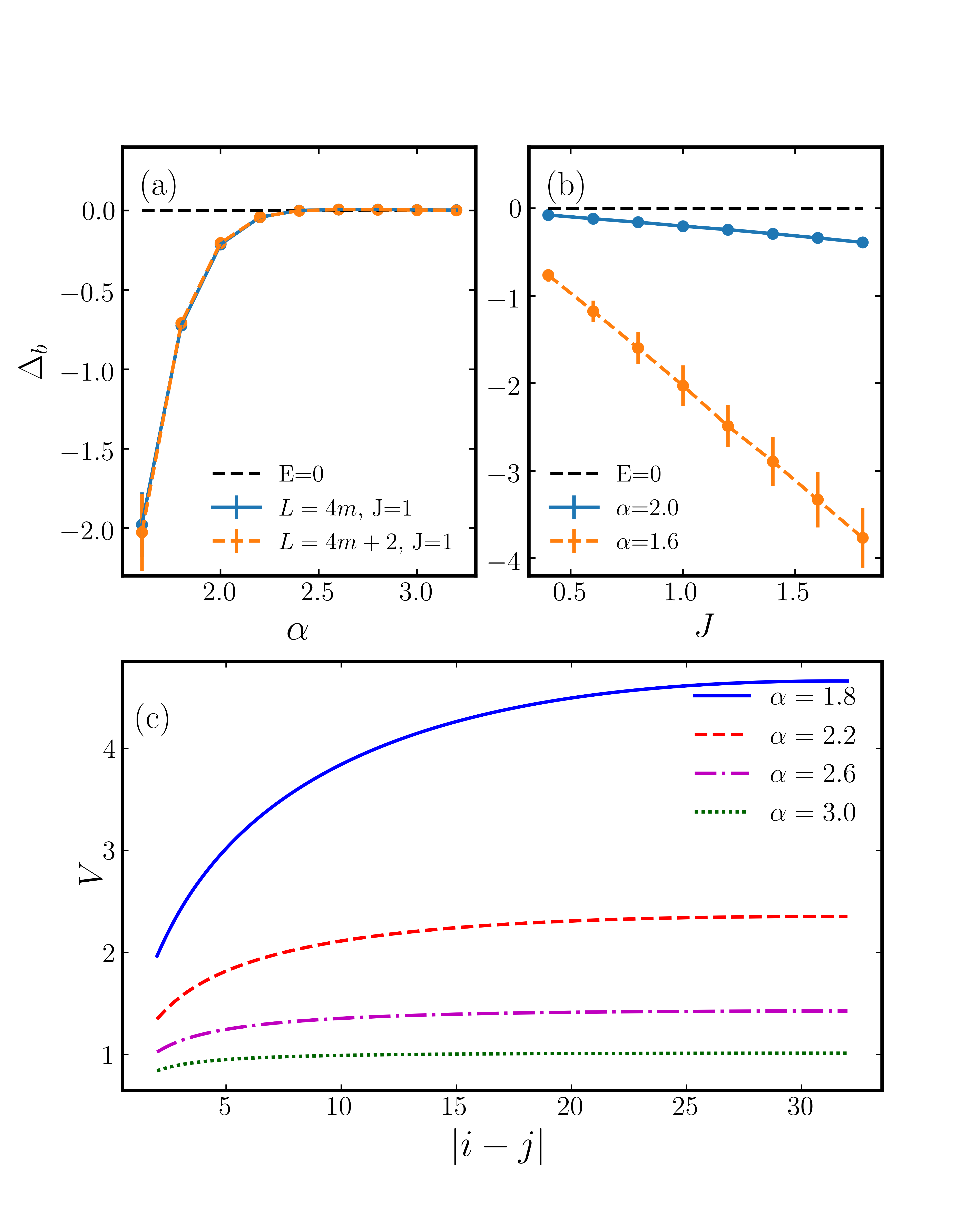}
	\caption{Binding energy between holon and spinon as a function of (a) $\alpha$ for fixed $J=1$ and (b) as a function of $J$ for different values of $\alpha$, extrapolated to the thermodynamic limit using chains of lengths $L=4m$ and $L=4m+2$ (see text). (c) Confining potential between holon and spinon in the Ising limit.}
	\label{fig:binding_confining}
\end{figure}

In Figs.\ref{fig:phase_diagram}(b) and (c) we show similar phase diagrams for the $t-J$ model with long-range RKKY interactions and for two values of exponent $\alpha=1.6,2.0$. We firstly recall that at half-filling the system undergoes a transition from spin liquid for large $\alpha$ to N\'eel AFM from small $\alpha$ at a value of $\alpha_c \sim 2.2$. The long-range interaction plays the role of enhancing the AFM order. As a consequence, upon doping with holes, electrons tend to clump together in a large ordered domain, displacing the holes toward the boundaries of the chain. Therefore, smaller alpha translates into a growing phase separated region that dominates the phase diagram, also pushing the metallic phase to lower densities. On the other hand, at a value of $\alpha=2$ close to the transition we see that the $\Delta N=2$ regime survives albeit in a much narrower window. However, pairing does not survive. To show this, we calculate the pair binding energy describing the energy gain for creating a pair of holes
\begin{equation}
\begin{split}
    & E_b = (E_{2holes} - E_0) - 2(E_{1hole} - E_0) \\
    & \ \ \ = E_0 + E_{2holes} - 2E_{1hole},
\end{split}
\end{equation}
where we have defined $E_{2holes}=E(N=L-2,S^z=0)$; $E_{1hole}=E(N=L-1, S^z=1/2)$ and $E_0=E(N=L,S^z=0)$. Finite size extrapolations of $\Delta_b$ and the spin gap $\Delta_s$ are shown in Fig.\ref{fig:spin_gap} clearly indicate that both are zero in the thermodynamic limit within our error bars.

To offer more insight into this issue, we also calculate the spin-spin correlation:
\begin{equation}
S(r) = \langle S^z_0S^z_r\rangle, 
\label{Sr}
\end{equation}
density-density correlation: 
\begin{equation}
N(r) = \langle n_0n_r\rangle - \langle n_0\rangle \langle n_r\rangle
\label{Nr}
\end{equation}
and pair-pair correlation:
\begin{equation}
P(r) = \langle \Delta^\dagger_0\Delta_r\rangle 
\label{Pr}
\end{equation}
where $\Delta^\dagger$ operator represents creation operator for a singlet pair on neighboring sites:
\begin{equation}
\Delta^\dagger_i = \frac{1}{\sqrt{2}}(c^\dagger_{i,\downarrow}c^\dagger_{i+1,\uparrow} - c^\dagger_{i,\uparrow}c^\dagger_{i+1,\downarrow})
\end{equation}
In Fig.\ref{fig:correlations_4figs} we compare the long distance behavior of these correlations for various $J$ values, and we find no indication of dominant pairing, in agreement with the previous considerations. Therefore, we are led to conclude that the $\Delta N=2$ regime is due to a mismatch between the spin and charge velocities. As we will discuss below, for decreasing $\alpha$ holes and spinons will bind into composite quasi-particles, leading to the $\Delta N=2$ window to completely disappear. According to these observations, we deduce that long range antiferromagnetic interactions tend to destroy pairing in favor of phase separation, even at low densities. 

We finally comment on the small steps appearing at densities $n \sim 1$ in Fig.\ref{fig:N_mu}. The first step from the top shows that the single hole configuration is energetically robust. This is a singular case, since it is a $1/L$ effect and speaking of phase separation with only one hole has no significance. However, in open chains we find that the first few holes may tend to cluster at the edges of the chains. 

\begin{figure}
	\centering
	\includegraphics[width=0.45\textwidth]{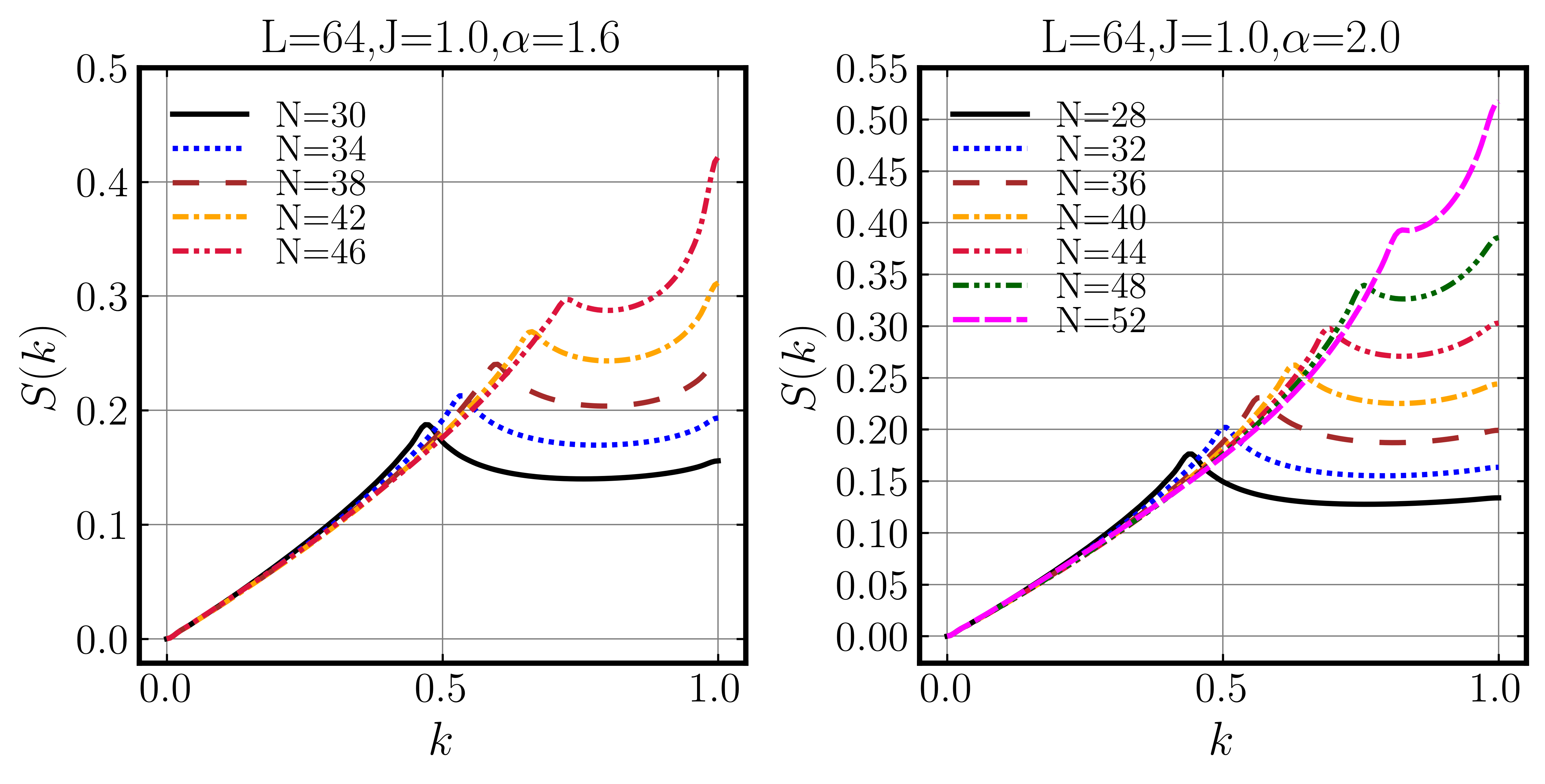}
	\caption{Spin-spin correlation in momentum space for $J=1$ and (a) $\alpha=1.6$ and (b) $\alpha=2$ and several densities.}
	\label{fig:sk}
\end{figure}

\section{Quasi-particle regime}
\label{sec:polaron}


As a consequence of the attraction between spinons and holons, it is possible to realize composite fermionic states that propagate coherently and carry both, spin and charge, degrees of freedom. These states have been referred-to in the literature as ``polarons'', ``mesons'' or ``string states'' (see {\it e.g.} Ref.\onlinecite{Bohrdt2020} for a discussion). In this work we favor the idea of a composite state in which spinon and holon ``orbit'' each other as in a Rydberg-like atom or diatomic molecule where the potential that holds them together, or glue, is a consequence of the long range antiferromagnetic interactions. The argument in favor of quasi-particle formation is more easily understood when presented in the limit of Ising-like interactions in the $t-J_z$ model\cite{Chernyshev1999,Chernyshev2000,Batista2000,Chernyshev2002,Chernyshev2003,Chernyshev2007,Smakov2007,Bieniasz2019,Wrzosek2021,Grusdt2018parton,Grudst2020}: as a hole is created in the antiferromagnetic background, it does so accompanied by a domain wall, a spinon (see Fig.\ref{fig:polaron}). A contact-like local potential proportional to $J$ tends to bind them, but is easily overcome by the kinetic energy of the hole. However, in the presence of RKKY interactions, the effective binding potential is non-local and grows sub-linearly with the separation distance $r$ between hole and spinon due to the string of unaligned spins left behind, as shown in Fig.\ref{fig:polaron}(c): 
\begin{equation}
V(r) = E_{Ising}(r) - E_{Ising}(r=1),
\label{potential}
\end{equation}
\begin{equation}
E_{Ising}(r) = \sum_{i\neq j}(-1)^{i-j+1}\frac{\langle 0,r|S_{i}^zS_{j}^z|0,r\rangle}{|i-j|^\alpha},
\end{equation}
where the state $|i,j\rangle$ represents a holon at position $i$ and a spinon and position $j$. 
In Fig. \ref{fig:binding_confining}(c) we show the profile of this potential for different values of $\alpha$ obtained in the Ising limit. Due to the long-range nature of the interactions, the corresponding energy cost would grow with the number of anti-aligned spins. As a result, holon and spinon will now energetically prefer to stick together as a composite object. 


In the fully $SU(2)$ spin rotational case, we can numerically calculate the binding energy $\Delta_b$ between holon and spinon following a prescription proposed in Ref.\cite{Smakov2007,Chernyshev2007}. This quantity can be obtained as:
\begin{eqnarray}
\Delta(L) & = & E_p(L) - E_s(L) - E_h(L) \\
& = & [E(L,L,0) + E(L,L-1,1/2)]  \nonumber \\
& - &\cfrac{1}{2}[E(L-1,L-1,1/2) + E(L+1,L+1,1/2)] \nonumber \\
& - & \cfrac{1}{2}[E(L-1,L-2,0) + E(L+1,L,0)] \nonumber
\end{eqnarray}
where $E(L,N,S^z)$ represents the ground state energy of a system with length $L$, particle number $N$ and total spin $S^z$, and $L$ is taken to be even in this definition.  The spinon energy $E_s$ is determined by the average ground state of chains with $L\pm 1$ sites at half-filling; the holon energy $E_h$ is obtained by adding one hole; finally, the spinon-holon quasi-particle energy $E_p$ is given by the ground state of chains with $L$ sites at half filling with and without one hole. In these calculations we used periodic boundary conditions with 1600 DMRG states to keep the truncation error under $10^{-6}$ for system sizes up to $L=44$.
In order to obtain a better extrapolation to the thermodynamic limit we divided the calculation into two groups using (i) $L=4m$ and $L=4m-1$ (ii)  $L=4m+2$ and $L=4m+1$, with $m \in \mathbb{Z}$. To make the extrapolation better conditioned, we flip the sign of hopping term to transfer the lowest energy from $k=\pi$ to $k=0$ for the chains with length $4m+1$ and one hole(see Ref.\onlinecite{Chernyshev2007} for details). The extrapolated results as a function of $\alpha$ are shown in panel (a) of Fig.~\ref{fig:binding_confining}. When $\alpha$ is increased through the antiferromagnetic transition into the spin-liquid phase, the binding energy vanishes. We also notice that the data for both $L=4m$ and $L=4m+2$ have consistent extrapolations, so we only include the results for $L=4m+2$ sector in panel (b) of Fig.~\ref{fig:binding_confining}, which shows the dependence with $J$ for different values of $\alpha$. Our results indicate a dramatic increase (practically exponential) of the binding energy upon entering into the antiferromagnetically ordered phase.

Notice that similar arguments can be used to explain pairing near half-filling, since the same confining potential would also act between two holes. However, in the presence of RKKY interactions this potential is so strong that forces the holes to clump together and the system to phase separate, as observed to occur in the phase diagram near half-filling. 

When the system is deep in metallic phase, a spin density wave instability appears as a cusp at $k=2k_F$ in the spin correlations in momentum space:
\begin{equation}
S(k) = \cfrac{1}{L}\sum_{i,j} e^{ik(i-j)} \langle S^z_iS^z_j\rangle. 
\label{sk}
\end{equation}

\begin{figure}
	\centering
	\includegraphics[width=0.45\textwidth]{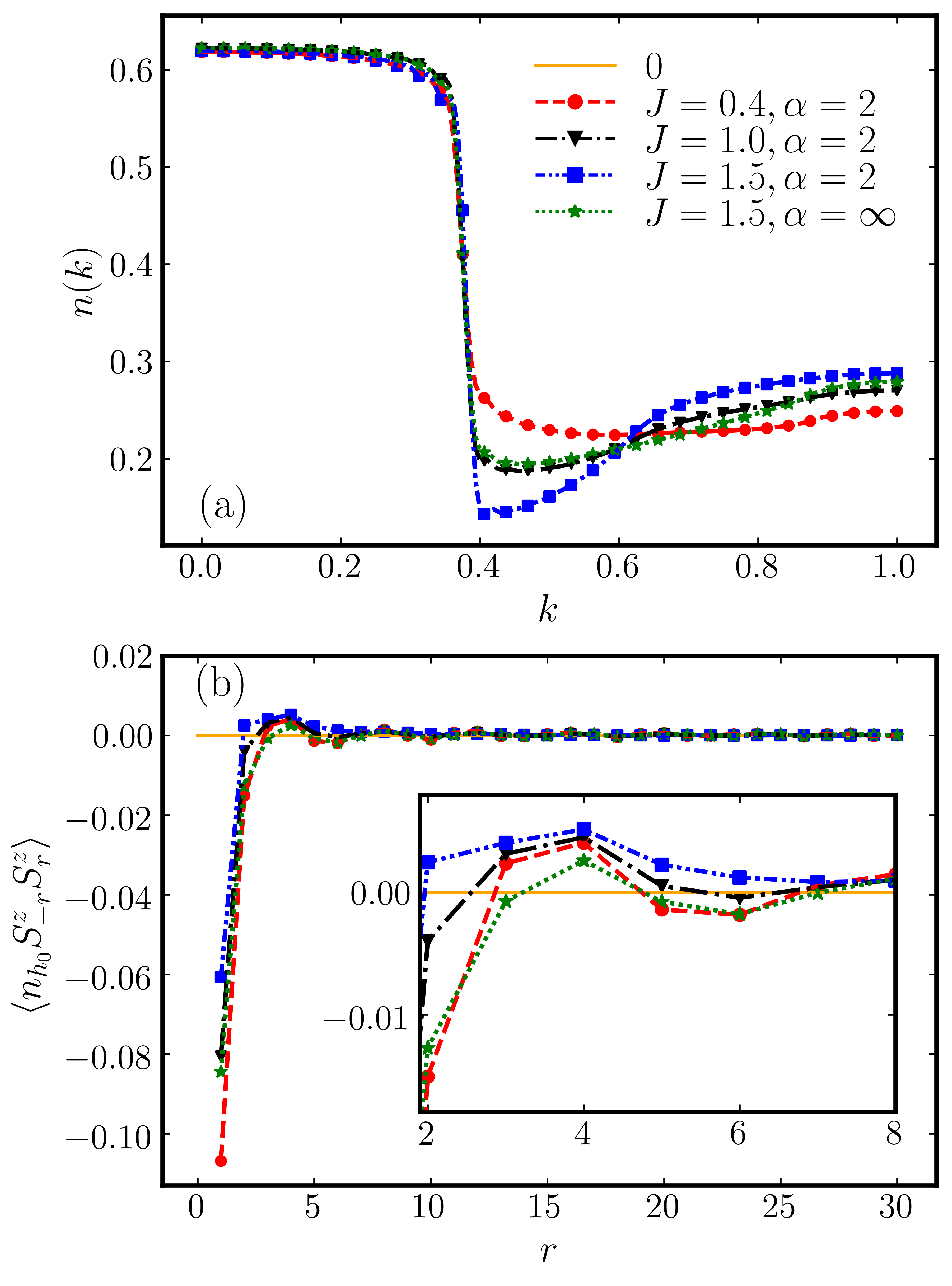}
	\caption{(a): Momentum distribution function (b): Spin-spin correlations across the hole as defined in the text. The inset shows an enlarged part of the main figure. Results are for a chain of length $L=64$, density $n=0.75$, and parameters in the legend.}
	\label{fig:nk}
\end{figure}
In Fig.\ref{fig:sk} we show the spin structure factor for several values of $J$ and total density $n=N/L$. One can see that as the density increases, a dominant peak appears at $k=\pi$, induced by the RKKY exchange term. The boundary between the low density metallic phase and the high density regime with antiferromagnetic spin correlations is demarked by a dot-dashed line in Figs.\ref{fig:phase_diagram}(b) and (c). As observed here, this double-peak structure is not related to phase separation. We postulate that in this window of the phase diagram labeled as ``QP'', fermionic composite quasi-particles are stable, and that the AFM order survives {\it inside} the quasi-particles, which can have a large characteristic ``size''. 

As the density is lowered and we enter into the metallic phase, antiferromagnetic correlations are still dominant and the composite quasi-particles persist for a small range of parameters before spinon and holon finally deconfine. We offer two indirect indications that this is the case. 
The first evidence of fermionic composite states comes from the momentum distribution function $n(k)$, that displays a kink around $k=k_F$ right after crossing the boundary in the phase diagram, as seen in Fig.\ref{fig:nk}(a). This implies the possibility of a jump or discontinuity, instead of a singularity, a sign of finite quasi-particle weight (unfortunately numerical uncertainty makes the calculation of this quantity very unreliable). We also define the correlations across the hole\cite{Martins2000b,Hilker2017} as $\langle n_{h,0}S^z_{-r}S^z_{r}\rangle$, where the hole is projected on the reference site $0$ which is taken to be at the center of the chain. The results shown in Fig.\ref{fig:nk}(b) are normalized by $\langle n_{h,0} \rangle$. In the quasi-particle regime we find that spins equidistant from the hole are aligned in the same direction. This is consistent with a charge and spin configuration as the one depicted in Fig.\ref{fig:polaron}(a), corresponding to a composite state of a spinon and a holon. Notice that the correlations at distance $r=1$ in Fig.\ref{fig:nk}(b) are antiferromagnetic, indicating fluctuations with a heavy contribution from the configurations in Fig.\ref{fig:polaron}(b), that is to be expected since the quasi-particle moves combining hopping and spin-flip processes. Outside of the quasi-particle regime, the correlations across the hole tend to oscillate with momentum $2k_F$ indicating deconfined spinons and holons, as illustrated in Fig.\ref{fig:polaron}(c). The presence of coherent quasi-particles will be supported by calculations of the photoemission spectra in the following section. 

\begin{widetext}

\begin{figure}
	\centering
	\includegraphics[width=0.95\textwidth]{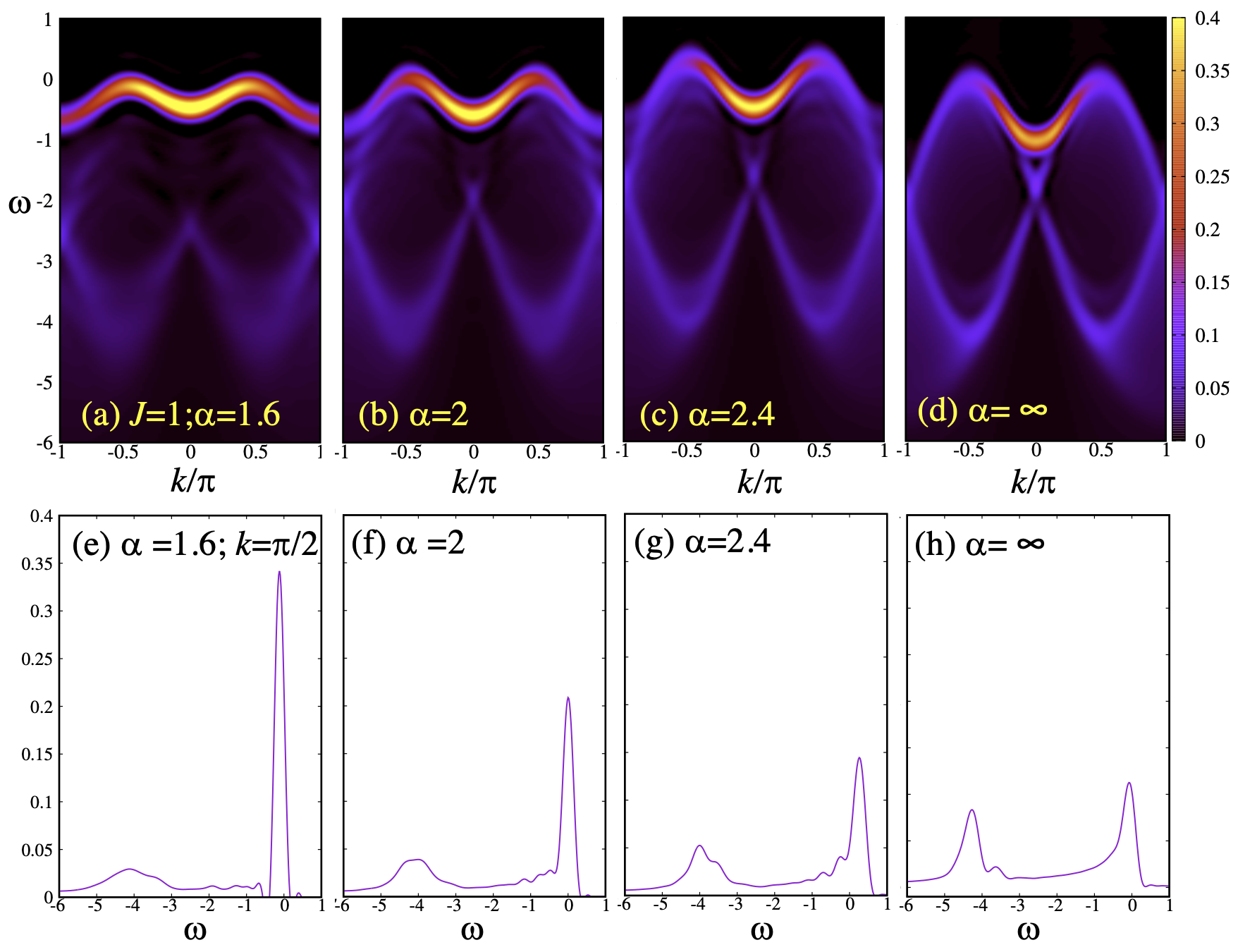}
	\caption{Photoemission spectrum at half-filling $n=1$ for $J=1$ and different values of $\alpha$ ($\alpha=\infty$ indicates the conventional $t-J$ model). The lower panels  display cuts in frequency along the $k=\pi/2$ line showing the development of a coherent quasi-particle peak. }
	\label{fig:spectrum_vs_alpha_n=1}
\end{figure}
\end{widetext}

\section{Single hole spectral function}
\label{sec:spectrum-hole}

\subsection{DMRG at half-filling}
We are seeking signatures of coherent quasi-particles in the photoemission spectrum of the generalized $t-J$ model with long-range RKKY interactions using the time-dependent DMRG method (tDMRG) \cite{white2004a,daley2004,vietri,Paeckel2019}. We follow the standard prescription detailed in the original work Ref.\cite{white2004a} and subsequent studies of the Hubbard model\cite{Feiguin2009,Nocera2018}. We calculate the two-time correlator:
\begin{equation}
\langle c^\dagger_{r\uparrow}(t)c_{0\uparrow}(0) \rangle = \langle \psi_0|e^{iHt}c^\dagger_{r\uparrow} e^{-iHt}c_{0\uparrow}|\psi_0 \rangle,
\end{equation}
where $c_{0\sigma}$ here is defined at the center of the chain, and $r$ is the distance from center. By Fourier transforming to momentum and frequency, we reconstruct the momentum resolved spectral function. This procedure is carried out numerically over a finite time window $t_{max}$ with $t_{max}=20$ unless otherwise stated. In order to attenuate artificial ringing we use standard windowing techniques. The spectrum will exhibit an artificial broadening that is inversely proportional to $t_{max}$. The long-range terms in the Hamiltonian make it convenient to use a time-step targeting procedure with a Krylov expansion of the time-evolution operator \cite{Feiguin2005} and a time step $\delta t=0.1$ (time is measured in units of hopping $t^{-1}$ and $t$ is our unit of energy). We study chains of length $L=48$ using up to 400 DMRG states that guarantees that the truncation error remains smaller than $10^{-6}$ over the time window. In all results shown here we introduced a shift in $\omega$ given by $\mu=E_1-E_0$, where $E_0$ is the energy of the ground state $|\psi_0\rangle$ with $N=L$ and $E_1=\langle \psi_1|H|\psi_1\rangle/\langle \psi_1|\psi_1\rangle$ with $|\psi_1\rangle=c_{L/2}|\psi_0\rangle$.

\begin{figure}
	\centering
	\includegraphics[width=0.45\textwidth]{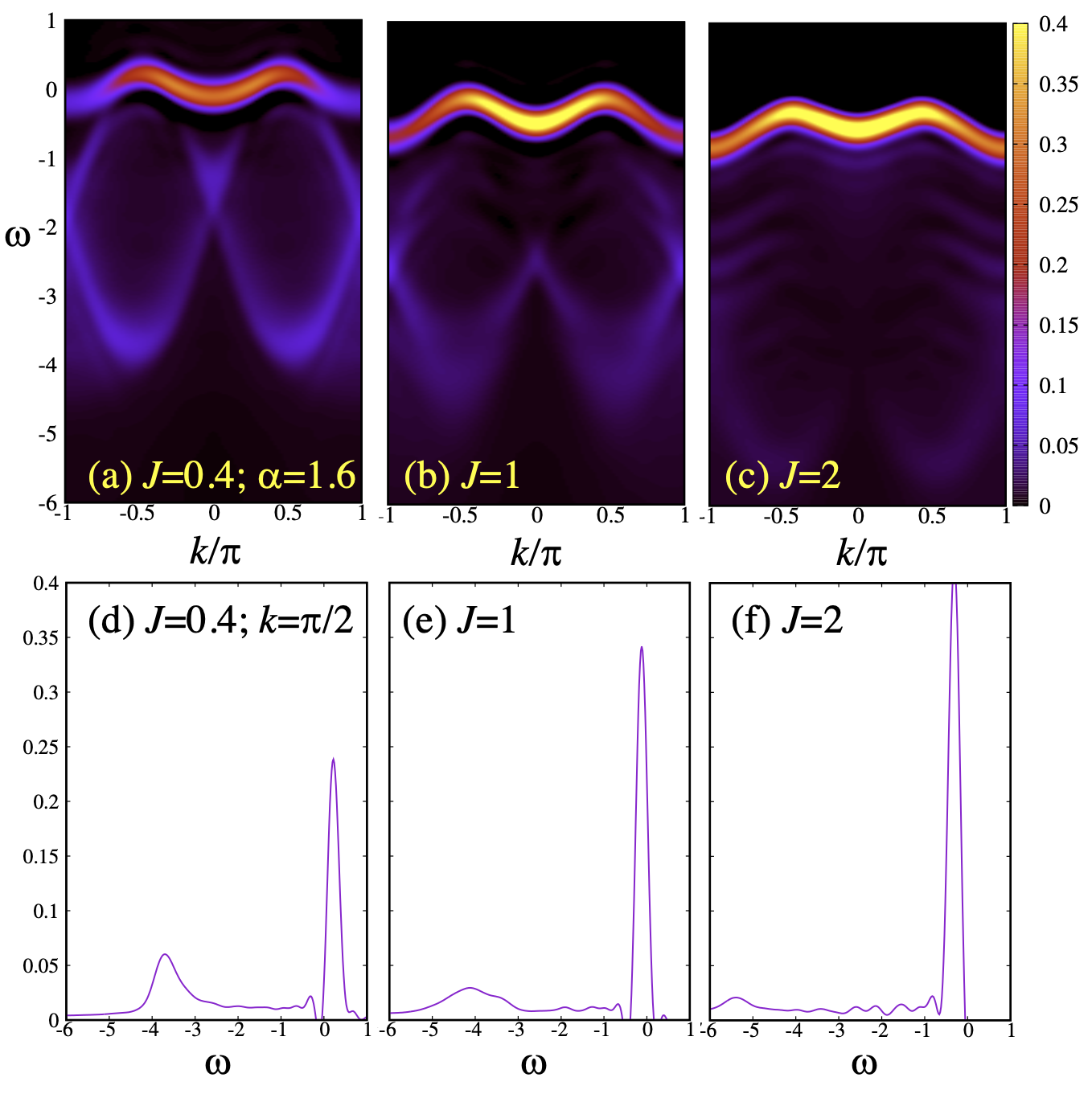}
	\caption{Top row: Photoemission spectrum at half-filling $n=1$ for $\alpha=1.6$, deep into the N\'eel phase and different values of $J$. Bottom row: cuts in frequency along the $k=\pi/2$ line showing the development of a coherent quasi-particle band splitting from the continuum. }
	\label{fig:spectrum_n=1}
\end{figure}

We show results at half-filling in Fig.\ref{fig:spectrum_vs_alpha_n=1} for $J/t=1$ and varying $\alpha$ across the transition from N\'eel to spin liquid. The spectral function of the conventional $t-J$ is displayed in panel Fig.\ref{fig:spectrum_vs_alpha_n=1}(d).  The spectrum displays features of both spinon and holon dispersions\cite{Suzuura1997,Brunner2000}: assuming holon and spinons dispersions $\epsilon_h(q_h)$ and $\epsilon_s(q_s)$, one can construct all possible energies with momentum $k$ as $\epsilon(k)=\epsilon_h(q_h) + \epsilon_s(q_s)$, with $k=q_h+q_s$. Clearly, this construction will yield a continuum of energies with momentum $k$. 
The figures show the development of a coherent quasi-particle peak as we cross the critical value of $\alpha_c \sim 2.2$ from above. 
In addition, we observe that the dispersion develops two minima. This is explained by realizing that the composite quasi-particles will have to move by means of a combination of hopping and spin flips. Therefore, the particle will effectively acquire a second (next-nearest) neighbor hopping contribution since each spin flip moves a spinon by two sites. 
In the lower panels of the same figure we show cuts along the $k=k_F=\pi/2$ line. We can clearly resolve the quasi-particle peak splitting from the upper edge of the continuum (ringing oscillations are artifacts of the Fourier transform, as noted above). 

\begin{figure}
	\centering
	\includegraphics[width=0.45\textwidth]{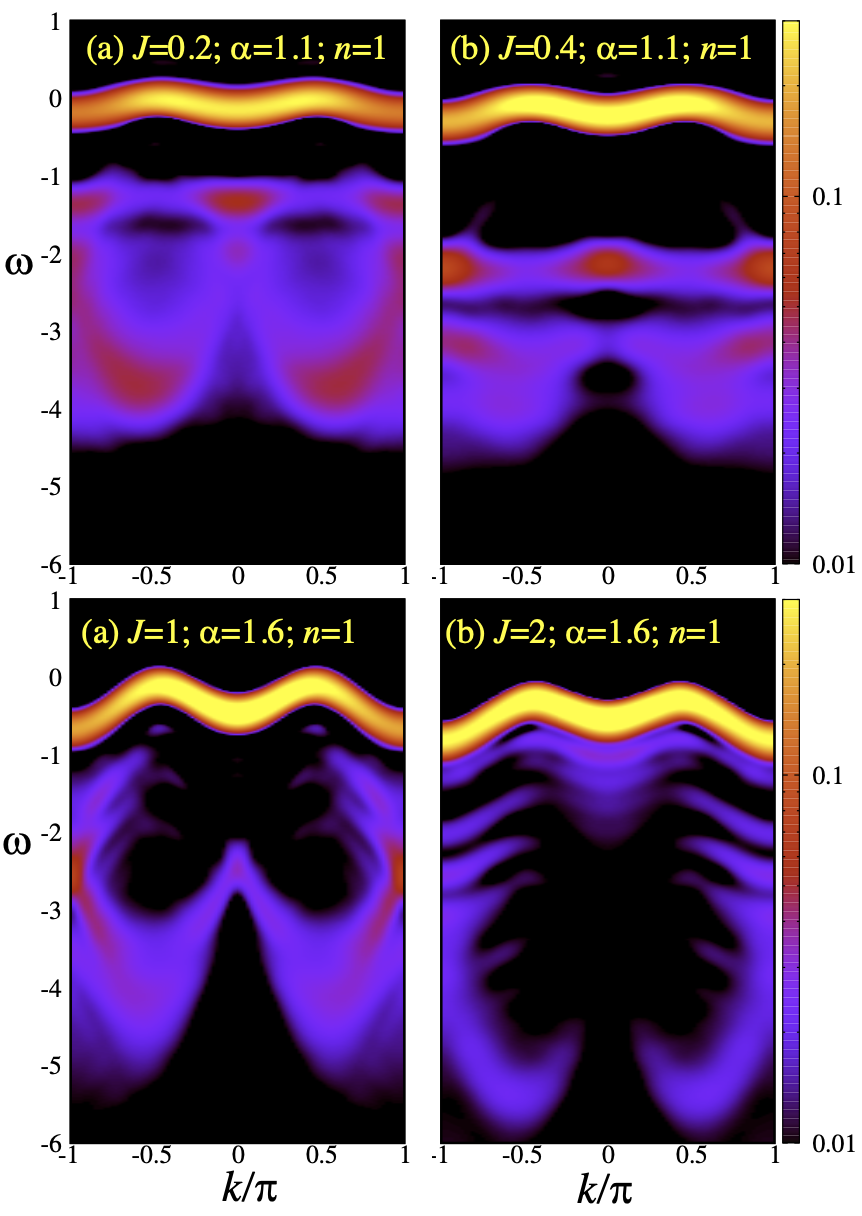}
	\caption{Photoemission spectrum for a single hole for several values of $J$ and $\alpha$ in a color log scale, showing features associated to strings inside the continuum.}
	\label{fig:spectrum_n=1_logscale}
\end{figure}

In Fig.\ref{fig:spectrum_n=1} we compare the behavior with varying $J$ at a fixed value of $\alpha=1.6$, deep into the AFM ordered phase, where we see an increase of spectral weight in the quasi-particle band with increasing $J$. Looking more carefully, we notice the development of new structures inside the incoherent continuum. In order to resolve these new features, we plot the spectral weight in a log scale in Fig.\ref{fig:spectrum_n=1_logscale}. The ``ladder'' appearing inside the continuum is not a numerical artifact but a manifestation of the string confining potential (\ref{potential}). These ``string'' excitations are not stable, and decay into a spinon and a holon as we discuss below. We also look at two extreme cases with $\alpha=1.1$ and small $J=0.2$ and $0.4$ in panels (a) and (b) of the same figure, where we observe just one or two prominent string states. One way to interpret the energy spacing between them is by considering a simplistic picture in which spinons cannot move and holes behave as particles trapped by the confining potential of Fig.\ref{fig:binding_confining}. In a linear potential, these bound states would be Airy functions with equally spaced energy levels \cite{Batista2000}. In our case, the behavior is less trivial, but analogous, with a spacing that increases with increasing $J$ or with smaller $\alpha$. We provide a more detailed theoretical description and analysis in the following sections. 

\begin{figure}
	\centering
	\includegraphics[width=0.45\textwidth]{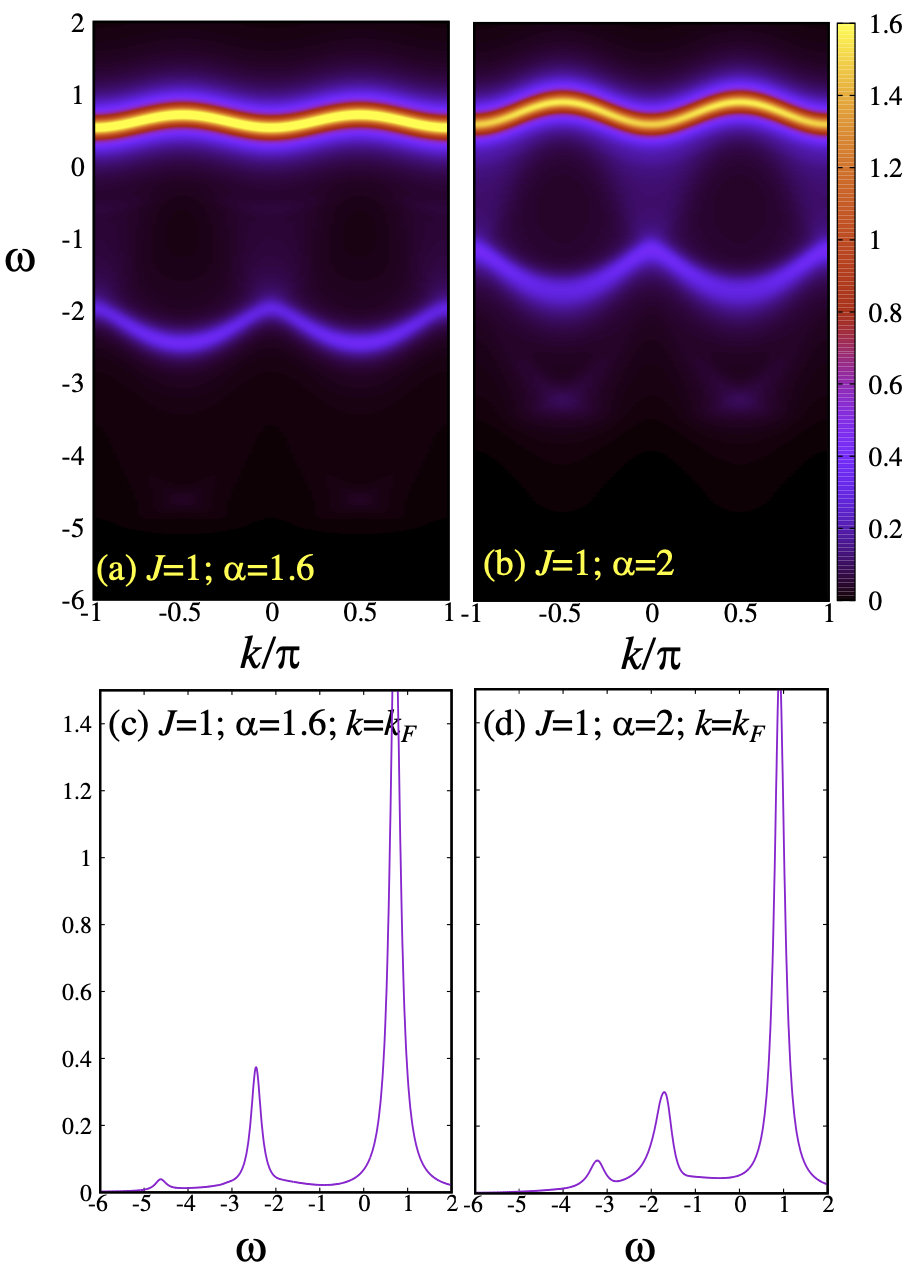}
	\caption{Photoemission spectrum at half-filling $n=1$ for $J=1$ and different values of $\alpha$ obtained with the self-consistent Born approximation (SCBA). In the lower panels (c) and (d) We show cuts in frequency along the $k=\pi/2$ line. }
	\label{fig:scba_J1}
\end{figure}

\subsection{SCBA}

In order to gain further physical insight, we compare the DMRG photoemission spectra with the predictions of an 
analytic approach that correctly describes the motion of a hole tightly coupled with the semiclassical spin-wave 
excitations of an antiferromagnetic state, giving rise to a spin-polaron quasiparticle.  
Hence, we have calculated the single hole spectral functions by means of the self-consistent Born approximation 
(SCBA)~\cite{Martinez1991, Liu1992, Horsch1994, Reiter1994}, a method that has been proven to compare quantitatively 
very well with exact diagonalization (ED) results on finite two-dimensional clusters with short range interactions in different 
antiferromagnets~\cite{Martinez1991, Lema97, Trumper04, Hamad2008, Hamad12}. 
It is one of the more reliable and checked analytical methods up to date to calculate the hole Green's function, 
and in particular, its QP dispersion relation. In order to do such calculation, we follow standard procedures \cite{Martinez1991}. 
On one hand, the magnetic elementary excitations are obtained treating the Heisenberg exchange terms of the Hamiltonian at the 
linear spin-wave (LSW) approximation. Thus, we restrict this description to the long range magnetically ordered regime of the 
phase diagram, whose magnetic spectrum consists of semiclassical magnons within the LSW approximation. 
In this sense, the SCBA will have the ability to exhibit the physics of a single hole interacting with one-dimensional 
magnons only, excluding other possible excitations like spinons. 
On the other hand, the electron creation and annihilation operators in the hopping terms are mapped into holons in the slave-fermion 
representation (details in the Supplementary Material of Ref.~\cite{Hamad2018}).  Within SCBA, we arrive to an effective Hamiltonian: 
\begin{equation}
H_{\text{eff}} =\sum_{\mathbf{k}}\omega _{\mathbf{k}}\theta _{\mathbf{k}%
}^{\dagger }\theta _{\mathbf{k}}+ \frac{2S}{\sqrt{N}} \sum_{\mathbf{kq}%
}\left(M_{\mathbf{kq}}h_{\mathbf{k}}^{\dagger }h_{\mathbf{k}-\mathbf{q}}\theta _{%
\mathbf{q}}+\mathrm{H.c.}\right), 
\end{equation}
where there is no hole tight-binding-like free hopping term, since the ground state magnetic pattern consists in a 
180$^\circ$ antiferromagnetic N\'eel order. The magnon dispersion relation $\omega_{\bf{k}}$ is given by ~\cite{Yusuf2004, Laflorencie2005}: 
\begin{equation}
\omega_{\bf{k}} =\sqrt{\varepsilon_{\bf{k}}^{2}-g_{\bf{k}}^{2}},
\label{omega}
\end{equation}
\begin{eqnarray}
\varepsilon_{\bf{k}} & = & 2JS \sum_{n=1}^{L/2} \frac{1}{(2n-1)^{\alpha}} - 2JS \sum_{n=1}^{L/2}{\frac{cos(2nk)-1}{(2n)^{\alpha}} }\nonumber \\   
g_{\bf{k}}& = & -2JS \sum_{n=1}^{L/2} {\frac{cos[(2n-1)k]}{|2n-1|^{\alpha}} }, 
\label{magnetic}
\end{eqnarray} 
\noindent and $M_{\mathbf{kq}}$ is the vertex interaction that couples the hole with magnons:
\begin{equation}
M_{\mathbf{kq}} = 2t[\cos(k) u_{k-q} + \cos(q) v_{k-q}], 
\end{equation}
where $u_k$ and $v_k$ are the usual Bogoliubov coefficients. We have evaluated the sums in (\ref{magnetic}) up to 
where convergence is reached within a given tolerance. 
The self energy is calculated within the SCBA taking into account non-crossing diagrams only, which leads to the self consistent equation 
\begin{equation}
  \Sigma_k(\omega) = \frac{1}{L}\sum_{q}\frac{|M_{kq}|^2}{\omega + i\epsilon - \omega_{k-q} - \Sigma_{q}(\omega-\omega_{k-q})}
 \end{equation}
from which the hole spectral function is obtained. 

In Fig.~\ref{fig:scba_J1} the SCBA spectrum is shown. All the SCBA results are for $L=100$ sites, as it was checked that it already accurately describes the 
thermodynamic limit. 
Similar to the DMRG results, Fig. \ref{fig:spectrum_n=1}, the SCBA spectrum also shows a well defined quasiparticle band 
and a high energy continuum. However, in this case, the high energy spectra is clearly composed of strings. 
Strings are well known manifestations of chains of misaligned spins left behind by the hole as it hops \cite{Martinez1991, Liu1992}. 
As previously discussed, as the hole hops, misaligned spins are left behind, 
creating an energy potential that binds the hole, promoting its return to the original position. As in 1D there are no closed 
Trugman loops \cite{Kane1989}, the only option for the hole to ``cure'' the strings of wrongly aligned spins is to retreace its path. In this picture, the energy cost of moving the hole increases with distance, such that it is favorable for the hole to return to its original position by reabsorbing the magnetic excitations, in this case magnons, in reverse order 
of creation. These processes produce a non-crossing diagrams that are precisely what lies underneath the SCBA. However, the presence of spin-flip interactions in the Hamiltonian 
offers an alternative channel for the magnetic fluctuations to repair the misaligned spins, giving the hole the possibility of moving coherently. These processes are responsible for the QP peak in the spectrum. In the case of the $t-J_z$ Ising case, it has been shown that the ($k$-independent) spectrum consists of several strings \cite{Bieniasz2019}. 
In our model the same physics appears when $\alpha \simeq 1$, as the magnetic order becomes almost classical, with a very low probability of spin-flip
processes. 
As $\alpha$ increases and the long range order is weakened, the QP spectral weight decreases, and the high energy string continuum gains weight, but the SCBA picture remains essentially the same. 

\begin{figure}
	\centering
	\includegraphics[width=0.45\textwidth]{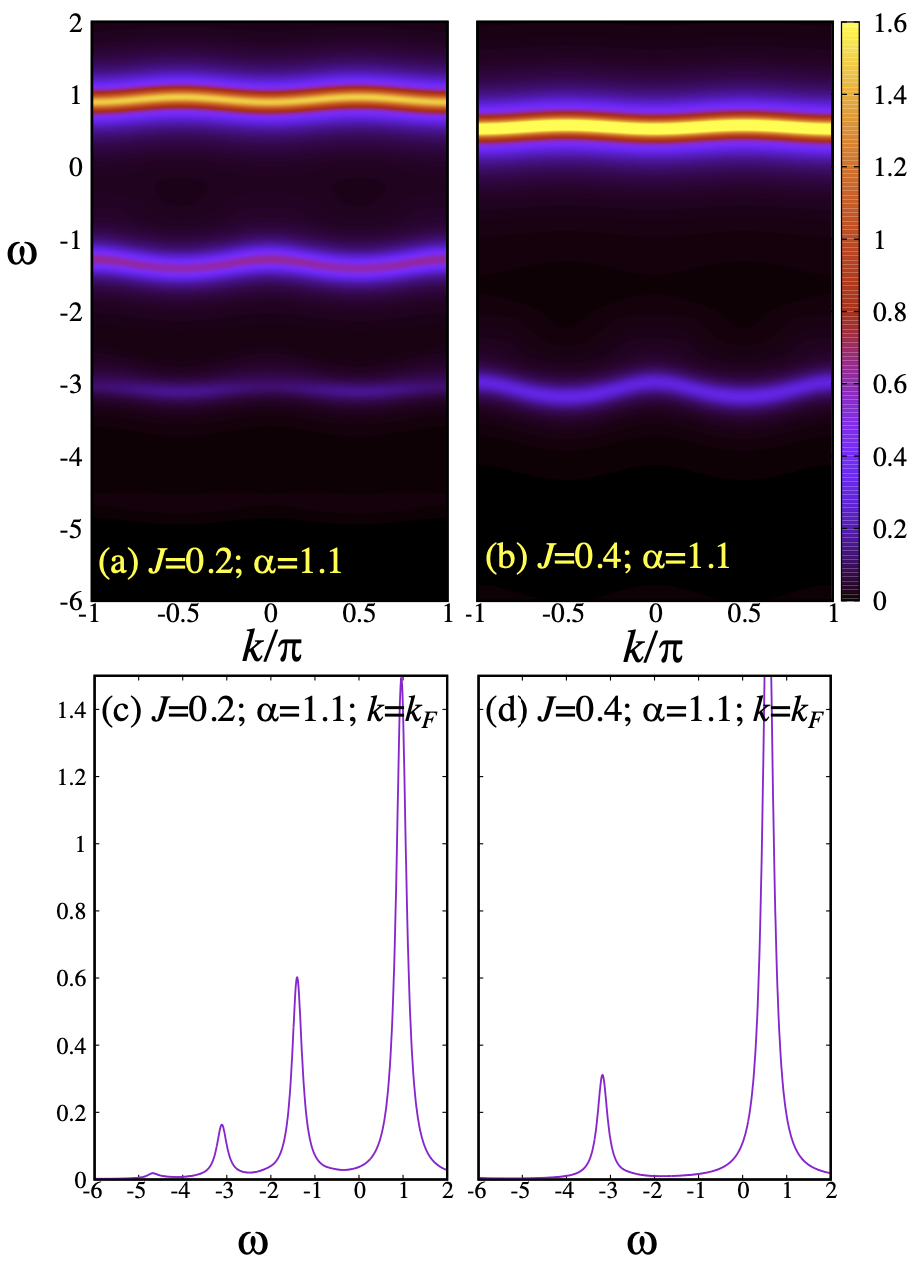}
	\caption{Photoemission spectrum at half-filling $n=1$ for $\alpha=1.1$ and different values of $J$ obtained with the self-consistent Born approximation (SCBA). In the lower panels (c) and (d) We show cuts in frequency along the $k=\pi/2$ line. }
	\label{fig:scba_alpha}
\end{figure}
 
In Fig.~\ref{fig:scba_alpha} we show the SCBA spectra at low $\alpha$, where the order is more rigid, varying $J/t$. At low $J/t$, where the characteristic time of magnetic fluctuations is much larger than that of the hole hopping, strings are manifested. 

On one hand, the QP spectrum of the SCBA and DMRG show good qualitative agreement in the ordered regime of the phase diagram, {\it i.e.} for $\alpha < 2.2$. This indicates that the QP in the ordered regime is effectively a magnetic polaron, where the hole dynamics is determined by its interaction with magnons. However, the high energy spectrum of the SCBA and DMRG differ, especially when $J/t$ and $\alpha$ are not very low. For low $\alpha$ both methods show a string picture. However, strings are unstable at high energies and decay into a continuum of spinon-holon excitations. On the other hand, in the SCBA approximation, strings spread over the entire energy range and are mostly responsible for the incoherent part of the spectrum. The noticeable differences between both methods in the intermediate to high energy sector arise because in the SCBA, due to the linear spin wave treatment of the magnetic spectrum, spinons are absent, and spin-charge separation is not possible. 

Hence, it can be concluded that the exact DMRG spectrum exhibits a spin-polaron quasiparticle at the lowest energy, and a few unstable strings for low $\alpha$, a typical behavior in higher dimension antiferromagnets. This QP is the result of the confinement of the holon and spinon excitations at low energies, while for higher energies there are signatures of spin-charge separation even for $\alpha < 2.2$, where there is long range order. Hence, the 1D RKKY system displays signatures of a dimensional crossover as a function of the energy, with the 1D physics surviving at high energies.



\subsection{Spinon-holon problem}

In order to provide an intuitive physical picture that accounts for spin charge separation and can allow us to peak into the internal structure of the composite quasi-particle, we start from the Ising limit, in which the ground state without a hole is just a trivial classical N\'eel order. When the insulator is doped with a hole it introduces a domain wall (a spinon) in the AFM background, as shown in Fig.\ref{fig:polaron}(a). Besides the motion of the hole, we consider additional quantum fluctuations mediated by spin-flip processes that allow the domain wall to move, ignoring processes that create new spinons for being energetically too costly (this includes long-range spin-flips). 

In order to make this scenario more concrete, we explicitly solve the two-body problem of a spinon and a holon. As noted previously, the spinon propagates by {\it two} sites with each spin-flip, and therefore it has a dispersion $\epsilon_s(k)=J\cos{(2k)}$, while the holon dispersion is $\epsilon_h(k)=-2t\cos{(k)}$. Both particle interact via a confining potential $V(r)$, Eq.\ref{potential}), where $r$ is the separation between the two. 
The formulation we use to study the two particle bound state been extensively applied in a number of scenarios in the literature, including Hubbard-like models\cite{Valiente2008,Valiente2009,Nguenang2009,Qin2008,Degli2014,Rausch2016,Rausch2017}, the formation of excitons in multi-band problems\cite{Yang2018}, and magnons \cite{Yang2020deconfined}. 
In our case, as the hole and spinon move apart, they leave behind a string of anti-aligned spins in the antiferromagnetic background that, unlike the conventional 1D $t-J$ model, costs an energy that grows with the relative distance between spinon and holon -- the length of the string. In our case, we assume the following Hamiltonian:

\begin{figure}
	\centering
	\includegraphics[width=0.5\textwidth]{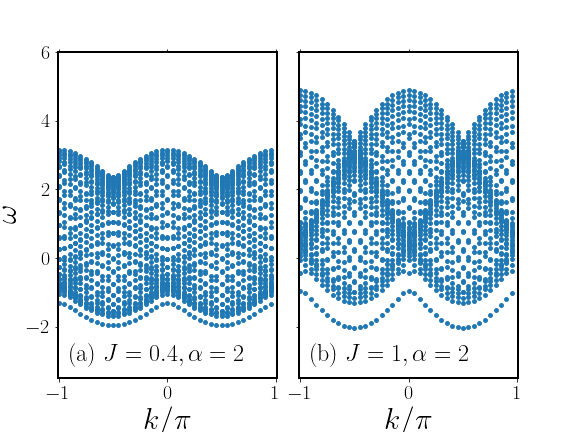}
	\caption{Spectrum of the spinon-holon problem for two different values of $J$ and $\alpha=2$. }
	\label{fig:strings}
\end{figure}

\begin{eqnarray}
H|r_s,r_h\rangle &=& -t (|r_s,r_h+1\rangle+|r_s,r_h-1\rangle) \nonumber \\
&+& J/2(|r_s+2,r_h\rangle+|r_s-2,r_h\rangle). \nonumber \\
&+& V(|r_s-r_h|)|r_s,r_h\rangle,
\end{eqnarray}
where $r_s$ and $r_h$ refer to the position of the spinon and holon respectively, and the potential $V$ is given by Eq.~(\ref{potential}).
We consider periodic boundary conditions, which allows us to construct a basis of states that are translational invariant and labeled by a momentum $k$:
\begin{eqnarray}
|r,k\rangle &=& \frac{1}{\sqrt{L}}\sum_{x=0}^{L-1} e^{ikx} T_x |r_s=0,r_h=r\rangle. \\ 
& = & \frac{1}{\sqrt{L}}\sum_{x=0}^{L-1} e^{ikx} |r_s=x,r_h=r+x\rangle \nonumber
\end{eqnarray}
In this basis, the Hamiltonian matrix can be easily obtained and numerically diagonalized for each momentum sector. The spectrum of a chain with $L=40$ sites is shown in Fig.\ref{fig:strings} for $\alpha=2$, $J=0.4$ and $1$. Without interactions, the spectrum consists of a continuum with a lower edge given by $\omega(k)=\epsilon_s(k)-\epsilon_h(k=0)=\epsilon_s(k)-2$. Long-range interactions favor the formation of composite fermionic bound states splitting from the spinon-holon continuum, resembling our numerical results obtained with DMRG and the SCBA in previous sections. However, in this picture the spinon-holon continuum is manifest while the SCBA cannot account for it. Despite being a crude approximation, it offers intuition about the nature of the quasi-particle excitations:
the hole and spinon form a Rydberg-like state or diatomic molecule, confined by the string potential of Fig.\ref{fig:binding_confining}. In order to move coherently, the spinon-hole pair has to do it in two steps: first, a spin flip moves the spinon by two sites; second, the holon needs to follow and settle in between the two parallel spins. As a consequence, the dispersion of the ``polaron'' will display two minima, as already observed, with a bandwidth determined primarily by $J$.

\section{Photoemission at finite doping}
\label{sec:spectrum}

\begin{figure}
	\centering
	\includegraphics[width=0.45\textwidth]{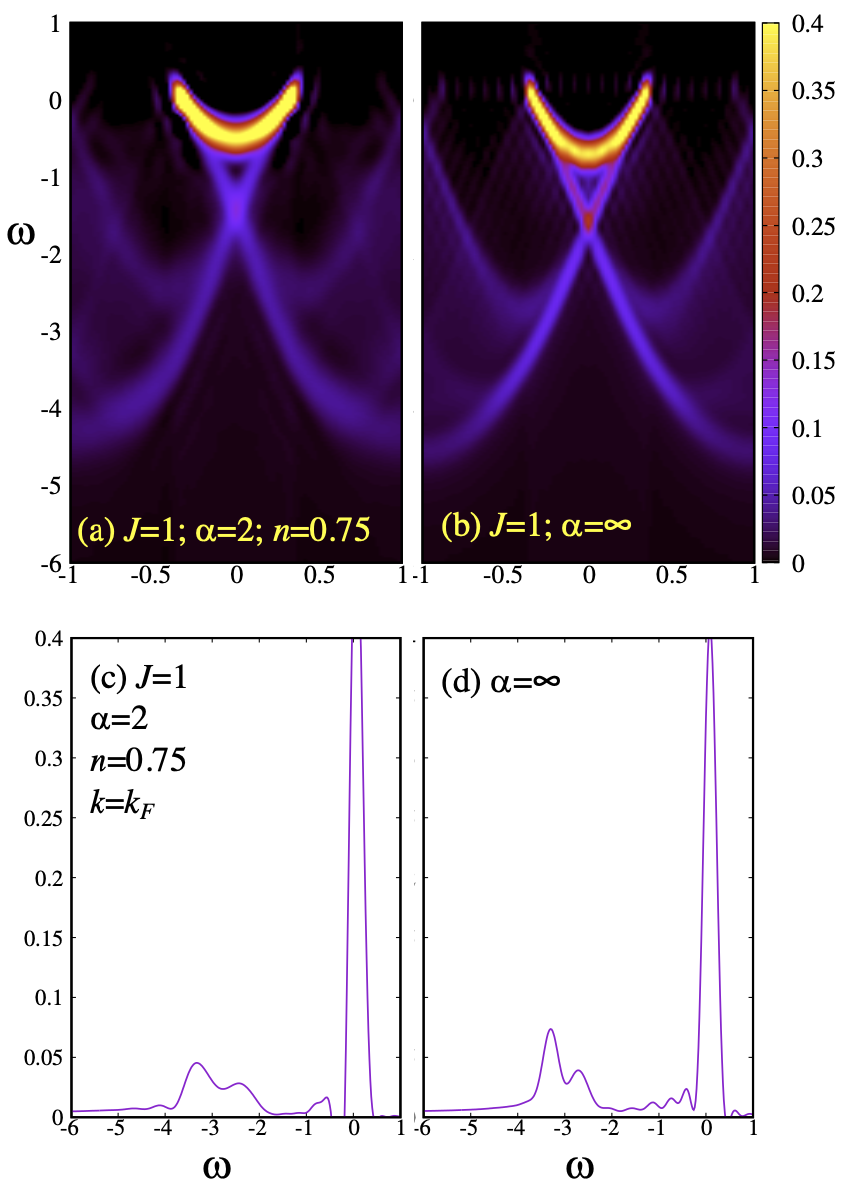}
	\caption{Photoemission spectrum at density $n=0.75$ for $J=1$, $\alpha=1.6$ and the conventional $t-J$ model ($\alpha=\infty$). We show cuts in frequency along the $k=k_F$ line showing the development of a coherent quasi-particle peak. }
	\label{fig:spectrum_n=075_J=1}
\end{figure}

At finite doping the existence of coherent quasiparticles in one-dimensional is less expected. One argument against it is the presence of pervasive nesting at all densities, making Fermi liquid theory unstable and LL theory apparently unavoidable. However, in Fig. \ref{fig:spectrum_n=075_J=1} we show the spectrum at momentum $k=k_F$ at density $n=0.75$ for the conventional $t-J$ model displaying a Fermi edge singularity, and $\alpha=2$, showing a coherent peak that seems to split from the continuum. This numerical evidence suggests that indeed quasiparticles may be stable, at least in a range of (relatively high) densities. We postulate, without offering a proof, that this occurs whenever the spin correlations have a dominant AFM peak at $k=\pi$, the region labeled as ``QP'' in Fig.\ref{fig:phase_diagram}. 


\section{Conclusions}
We have studied the stability of fermionic quasiparticles in a doped antiferromagnet in a relatively simple one-dimensional model that realizes much of the phenomenology of the higher dimensional $t-J$ model. At half-filling, the all-to-all-interactions lead to a transition to a phase with spontaneous symmetry breaking and N\'eel order. Interestingly, pairing is no longer stable and gets replaced by a large region where phase separation into AFM and hole-rich domains takes place: long-range interactions lead to holes clumping, or clustering, indicating that in order to stabilize and have mobile pairs, a weaker confining potential might be required. Such scenarios have been explored in the conventional $t-J$ model with a staggered magnetic field where the potential is linear and pairing is robust\cite{Batista2000}, or in a square lattice with long range AFM order but where the holes are allowed to move only along the $x$ direction \cite{Grusdt2018scipost}. In the context of our model, one possible way to counteract the instability toward phase separation is by including a long range Coulomb repulsion and second-neighbor hopping to increase the hole kinetic energy. Work in this direction is underway.

Upon doping the antiferromagnet with a hole, we observe spinons and holons binding to form composite quasiparticles in the regime with long-range antiferromagnetism for small $\alpha$, while they remain deconfined in the spin-liquid phase for $\alpha > \alpha_c \sim 2.2$. These excitations appear in the photoemission spectrum in the form of a coherent band splitting from the edge of the continuum of width determined by the exchange $J$ and the exponent $\alpha$. This band has also been observed in calculations on the 2D Hubbard model\cite{Dahnken2004,Groeber2000,Kohno2010,Yang2016cpt,Wang2015} and $t-J$ model \cite{Bohrdt2020}. The composite nature of the quasi-particles is supported by calculations of the spinon-holon binding energy that show a dramatic enhancement upon transitioning to the N\'eel phase and it is analogous to the observed physics in doped two dimensional antiferromagnets\cite{Martinez1991,Hamad2008,Hamad2021}.
This picture is further confirmed by SCBA calculations and the spectra of the spinon-holon problem. While the system exhibits well defined fermionic quasi-particles, their internal structure can be described as a spinon and holon oscillating around a common center of mass. Also, the SCBA calculations exhibit high energy strings in the long range ordered regime, but this physics is present in the DMRG calculations only at very low $\alpha$, where the magnetic order is very rigid. This leads us to conclude that a single hole, even in the ordered phase, couples at high energies to magnetic excitations that are spinons instead of magnons.  

The physical size of the composite quasi-particle can be quite large, and will be dictated by both $J$ and $\alpha$ and will diverge at the transition point $\alpha_c$. As a matter of fact, it might be possible that close enough to $\alpha_c$ the size of the ``polaron'' can be larger than the chain length, in which case one would only see spin-charge separation. 
At higher energies, both spinon and holon deconfine and we observe a continuum that can be associated to the original dispersions of the two objects. Therefore, while the system exhibits higher-dimensional physics at low energies, the 1D physics of spin-charge separation re-emerges at higher energies. This implies that at finite temperature, larger than the binding energy between holon and spinon, the quasi-particles would decay into its original constituents, establishing a limitation to our experimental ability to resolve them. 

At finite doping, upon crossing the phase separated region, we encounter evidence of surviving quasiparticles near the Fermi points. This is a quite puzzling surprise, since one would expect a 1D metal to be a Luttinger Liquid due to nesting and the fact that the Fermi surface consists only of discrete points (we are loosely referring to the regime with spin-charge separation as the LL phase, even though the excitation spectrum is no longer linear in the presence of long-range interactions). In fact, at low densities, the AFM order melts and spin charge separation re-emerges. However, it may seem as though the confining potential is strong enough to induce dominant AFM interactions and coherent quasi-particles even away from half-filling, albeit in a narrow window of densities. 
We point out that one-dimensional metals with fermionic quasi-particles are indeed possible, but this typically occurs in gapped systems, such as ladders.\cite{Lin1998,Tian2021,White2015,Yang2019ladder} In these systems the spin and charge gap may survive at finite doping\cite{Konik2000}. However, our model Eq.~(\ref{hami}) is gapless in both channels. Further numerical and theoretical work is needed to elucidate the mechanisms that might possibly stabilize fermionic quasi-particles in this regime. 

Finally, this work demonstrates the interesting phenomenology that can arise from the inclusion on long range interactions and, in particular, establishes Eq.~(\ref{hami}) as a rich toy model Hamiltonian to study higher dimensional physics with methods usually considered more amenable to one dimensional problems. 

\section{Acknowledgments}
We thank Fabian Essler and Anders Sandvik for stimulating discussions.
L.Y. and A.E.F. acknowledge support from the National Science Foundation under grants No. DMR-1807814 and DMR-2120501. I. H. and L.O.M. are supported by grant PIP2015 No. 364 of CONICET (Argentina).


%

\end{document}